\documentstyle[12pt,epsfig]{article}
\newcommand{\be}{\begin{equation}}
\newcommand{\ee}{\end{equation}}
\newcommand{\bear}{\begin{eqnarray}}
\newcommand{\ear}{\end{eqnarray}}
\newcommand{\la}{\langle}
\newcommand{\ra}{\rangle}
\newcommand{\slp}{\raise.15ex\hbox{$/$}\kern-.57em\hbox{$p$}}
\newcommand{\slv}{\raise.15ex\hbox{$/$}\kern-.57em\hbox{$v$}}
\newcommand{\slet}{\raise.15ex\hbox{$/$}\kern-.57em\hbox{$\eta$}}
\newcommand{\slB}{\raise.15ex\hbox{$/$}\kern-.57em\hbox{$B$}}
\newcommand{\slb}{\raise.15ex\hbox{$/$}\kern-.57em\hbox{$b$}}
\newcommand{\slW}{\raise.15ex\hbox{$/$}\kern-.57em\hbox{$W$}}

\begin{document}
{}
\vspace{1cm}
\begin{center}
{\LARGE \bf A Form Factor Model }\\
\vspace{.3cm}
{\LARGE \bf for}\\
\vspace{.3cm}
{\LARGE \bf Exclusive
$B$- and $D$-Decays}\\
\vspace{1.5cm}
{\LARGE Berthold Stech}\footnote{e-mail: B.Stech@thphys.uni-heidelberg.de}\\
\vspace{.5cm}
{\it Institut f\"ur Theoretische Physik, Universit\"at Heidelberg,\\
 Philosophenweg 16,\\
 D-69120 Heidelberg, Germany}
\end{center}

\vspace{1.5cm}
\baselineskip12pt
\begin{abstract}\noindent
An explicit model is presented which gives the momentum
transfer-dependent ratios of form factors of hadronic currents. For
the unknown Isgur-Wise function and its generalization for transitions
to light particles a simple phenomenological Ansatz is added. The
model allows a calculation of all form factors in terms of mass
parameters only. It is tested by comparison with experimental data,
QCD sum rules and lattice calculations.
\end{abstract}

\baselineskip14pt
The knowledge and understanding of the form factors of hadronic
currents is of decisive importance for the determination of the quark
mixing parameters. Matrix elements of hadronic currents also play an 
important role in the description of nonleptonic decays \cite{1}. In heavy-to-heavy transitions it has become
possible to extract these hadronic form factors from semileptonic
decay data with good precision and in an essentially model-independent
way \cite{kk}. For transitions to light particles, on the other hand,
there is no symmetry one can apply, and so far also insufficient
experimental information. Quark model calculations can be very
helpful for heavy-to-heavy as well as for heavy-to-light
transitions. Although strict theoretical error limits cannot
be given, they provide a vivid picture of what is going on
and give numerous testable predictions for quite different
processes. A quark model describing energetic transitions
must necessarily be a fully relativistic one 
\cite{2}. Relativistic
quark models have, however, notorious difficulties
connected with the relative time of the constituents, the
covariant wave functions, and the quark propagators in
the confinement region. In this aricle I will circumvent
these difficulties by concentrating on the peak of the
wave function overlap in the triangle graph and by assuming
simple properties of the (light) spectator particle.

The hadronic form factors for semileptonic decays are defined
as the Lorentz-invariant
functions arising in the covariant decomposition of matrix elements
of the type
\be\label{1}
\la F|(\bar q_f \gamma_\mu(1-\gamma_5)q_i)|I\ra.\ee
$I$ and $F$ stand for the decaying and the emitted particle
(or resonance) with masses $M_I$ and $M_F$, respectively.
I will discuss the decay of an initial pseudoscalar particle
into a $O^-$ or $1^-$ ($S$-wave) meson.

There
are two form factors $(F_0,F_1)$ describing the transition to a
pseudoscalar
particle  and four form factors $(V, A_0, A_1, A_2)$
governing transitions to $1^-$ states. For
radiative transitions described by Penguin diagrams the matrix elements
\be\label{2}
\la F|(\bar q_f\sigma_{\mu\nu} q^\nu(1+\gamma_5)q_i)|I\ra\ee
are needed. The decay to a $1^-$ state involves three form factors
$(T_1,T_2,T_3)$. The precise definition of the 9 form factors are 
given in Appendix A. A simplification occurs at zero momentum transfer
$q^2=0$:
\bear\label{3}
F_1(0)&=&F_0(0)\nonumber\\
A_0(0)&=&\frac{1}{2M_F}\left\{(M_I+M_F)A_1{(0)}-(M_I-M_F)A_2(0)\right\}
\nonumber\\
T_1(0)&=&T_2(0).\ear
In the case of heavy-to-heavy transitions, in the limit
in which the quarks active in the transition have infinite mass,
all nine form factors are given in terms of a single function
$\xi(y)$
called  the Isgur-Wise
form factor. The relations between the form factors arising in
this limit read
\bear\label{4}
&&F_1=V=A_0=A_2=T_1=T_3=\frac{1}{2}\frac{M_I+M_F}{\sqrt{M_I\cdot M_F}}
\xi_{\rm Isgur-Wise} (y)\nonumber\\
&&F_0=A_1=T_2=\frac{2\sqrt{M_IM_F}}{M_I+M_F}\frac{y+1}{2}\xi_
{\rm Isgur-Wise}(y)\ear
where $y=v_F\cdot v_I$ and $\xi_{\rm Isgur-Wise}(1)=1$.
 
In the realistic case of finite quark masses these relations
are modified;  each form factor depends separately on
the dynamics of the process.
Thus, eq. (\ref{4}) has to be generalized
by replacing for each form factor
 ${\cal F}$ the Isgur-Wise function by
\be\label{5}
\xi_{\rm Isgur-Wise}(y)\to h_{\cal F}(y)\xi_{FI}(y).\ee
One then has, for example,
\bear\label{6}
F_1&=&\frac{1}{2}\frac{M_I+M_F}{\sqrt{M_IM_F}}h_{F_1}(y)
\xi_{FI}(y)\nonumber\\
F_0&=&2\frac{\sqrt{M_F M_I}}{M_I+M_F}
\frac{y+1}{2}h_{F_0}(y)\xi_{FI}(y)\quad {\rm etc.}\ear
The function $\xi_{FI}(y)$ depends upon  the masses and properties
of the  initial and final state particles, but it does  not depend
on the  polarization or  the Dirac structure of the current.
In other words, this function is the same for all form factors
describing the transition $I\to F$. The functions $h_{\cal F}(y)$,
on the other hand, are different for each form factor. We choose them
in such a way that
\be\label{7}
h_{\cal F}(y)\to 1,\ \xi_{FI}(y)\to \xi_{\rm Isgur-Wise}(y)\ee
in the heavy quark limit. In the general case, i.e. for arbitrary
quark masses, we normalize  $h_{F_1}(y)$ and $h_{A_1}(y)$ in addition
to and consistent with (\ref{7}) at large values of $y$ where any
specific quark mass dependence of the transversal form factors
should die out:
\bear\label{8}
h_{F_1}(y\gg 1)&=&1\qquad {\rm for}\ 0^-\to 0^-\ {\rm transitions}
\nonumber\\
h_{A_1}(y\gg 1)&=&1\qquad {\rm for}\ 0^-\to 1^-\ {\rm transitions}.
\ear
\par
\begin{figure}[h]
\unitlength=1.0cm
\begin{picture}(15.,3.5)
\put(3.,0.){
\epsfxsize=8cm
 \epsffile{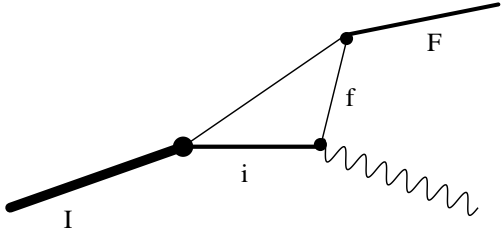}
}
\end{picture}
\caption{The triangle graph}
\end{figure}
\par
For a calculation of the functions $h_{\cal F}(y)$,
one has to consider the triangle graph of Fig. 1. It contains
unknown momentum-dependent couplings (the wave functions of
initial and final particles) as well as unknown quark propagators
in the confinement region. One may note, however,
that the integrand of
the transition amplitude as a function  of the spectator momentum
$p_{sp}$ is expected to have a sharp maximum where initial
and final wave functions overlap significantly. At this maximum the
spatial momentum of the spectator should vanish in a coordinate system
in which $\vec v_I=-\vec v_F$. Furthermore, at this maximum the energies
of initial and final quarks and the spectator should
have values close to
their masses in the rest system of the particles they belong to. This
will be the case for
\be\label{9}
\bar p_{sp}=\varepsilon_{sp}\frac{v_I+v_F}{y+1}\ee
where $\varepsilon_{sp}$ is an effective mass parameter for the (light)
spectator particle. Indeed, (\ref{9}) gives
\be\label{10}
v_I\cdot\bar p_{sp}=v_F\cdot\bar p_{sp}=\varepsilon_{sp}.\ee
For the initial $(i)$ and final $(f)$ quarks active in the process one
gets
\bear\label{11}
\bar p_i=M_I v_I-\bar p_{sp}\quad&,&\quad\bar p_f=M_F v_F-\bar p_{sp}
\nonumber\\
\bar p_i v_I=M_I-\varepsilon_{sp}\approx m_i\quad&,&\quad
\bar p_f v_F=M_F-\varepsilon_{sp}\approx m_f.\ear
Here the masses of initial and final quarks active in the
process are denoted
by $m_i$ and $m_f$, respectively. One also has from (\ref{9})
\bear\label{12}
\bar p^2_{sp}&=&\varepsilon_{sp}^2-\varepsilon^2_{sp}\frac{y-1}{y+1},
\nonumber\\
\bar p^2_i&=&(M_I-\varepsilon_{sp})^2-\varepsilon^2_{sp}
\frac{y-1}{y+1},\nonumber\\
\bar p^2_f&=&(M_F-\varepsilon_{sp})^2-\varepsilon^2_{sp}\frac{y-1}
{y+1}\ear
indicating small off-shell momentas even for large $y$ values.

Taking now the Dirac current and quark spin structure out of the
integral
for the transition amplitude by replacing $p_{sp}$ by $\bar p_{sp}$,
the functions $h_{\cal F}(y)$ can be obtained
from the covariant decomposition of the quantity
\be\label{13}
{\cal J}_\mu(\bar p_{sp})={\rm Tr}\Bigl\{\Gamma_\mu(\bar{\slp}_i
+m_i)
\gamma_5(m_{sp}-\bar{\slp}_{sp})(\gamma_5+\slet^*_F)
(\bar{\slp}_f+m_f)\Bigr\}\ee
in a straight forward way.
In (\ref{13}) the combination $\gamma_5+\slet_F$ represents the spin
wave function of pseudoscalar and vector particles in the final state.
$\Gamma_\mu$ stands for $\gamma_\mu(1-\gamma_5)$ or for $\sigma_{\mu\nu}
q^\nu(1+\gamma_5)$ in case of transitions described by Penguin diagrams.
The functions $h_{\cal F}(y)$ depend on mass ratios only and allow to 
predict the
ratios of form factors and thus in particular the polarization of the final
particle as a function of $y$ and the effective mass parameters $m_i, m_f, 
m_{sp}$ and $\epsilon_{sp}$. They are displayed in Appendix B. In the following $m_{sp}=\epsilon_{sp}$
is used in all applications.

The function $\xi_{FI}(y)$ is not calculable without a detailed
knowledge of the wave functions and the quark propagators in the
confinement region, which we are still lacking today. Therefore,
I shall make an Ansatz, which however respects scaling and
analyticity  requirements:
\bear\label{14}
\xi_{FI}(y)&=&\sqrt{\frac{2}{y+1}}
\left(\frac{1}{2}+\frac{1}{y+1}\right)
\nonumber\\
&&\left(1+\frac{y-1}{y+1}x_{FI}\right)^{-1}\cdot n_{FI}(y).\ear
The first factor is introduced because the form factors are assumed
to remain finite in the limit of vanishing mass of the final meson,
i.e., for
\be\label{15}
m_f,M_F\to 0,\ {\rm  keeping}\  q^2 \ {\rm and}
\  \frac{\varepsilon_{sp}}{M_F}\
{\rm fixed}\ee
implying $y\to\infty$.
With $\xi_{FI}(y)$ from (\ref{14}) this requirement is satisfied as
can be seen from (\ref{6}), the behaviour of $h_{\cal F}(y)$ and the
properties of $x_{FI}$ and $n_{FI}$ to be described below. The
second factor in (\ref{14}) is obtained from the infinite mass limit
of the quantity ${\cal J}(\bar p_{sp})$
in (\ref{13}) and was divided out when defining the functions
 $h_{\cal F}(y)$
according to (\ref{7}). The variable $\frac{y-1}{y+1}$ governs the off-shell
momenta according to (\ref{12}). The third factor in (\ref{14})
contains this variable and depends via the parameter $x_{FI}$
on the specific process considered. Because of the lack of spin
symmetry for light mesons this parameter is different, for instance,
for $B\to\pi$ and $B\to\rho$ transitions. One can fix it by requiring
$\xi_{FI}(y)$ to have a pole in $q^2$ at the position $M^*$ of the
nearest state or resonance carrying the quantum numbers of the weak
 current:
\be\label{16}
x_{FI}=\frac{y^*+1}{1-y^*}\ ,\ y^*=\frac{1}{2M_FM_I}
(M^2_I+M^2_F-M^{*2}).\ee
For decays to a pseudoscalar particle I will take the lowest $1^-$
state (the $B^*(5.325)$ for $B\to\pi$ transitions). For decays to vector
particles the lowest pole is caused by a $0^-$ state (the $B$-meson
pole in $B\to \rho$ transitions) which occurs in the $A_0$ form factor.

The last factor in Eq. (\ref{14}), $n_{FI}(y)$, is introduced to
normalize the form factors at $y=1$. One has to require
\be\label{17} h_{F_0}(1)\xi_{FI}(1)=1\ee
at least if particles $I$ and $F$ are identical.
Since the functions $h_{\cal F}(y)$ are normalized according to
(\ref{7}) and (\ref{8}), $h_{F_0}(1)$ is not equal
to 1 for general quark masses. $\xi_{FI}(y)$ has to correct for
that but should not spoil the independence on specific quark
mass differences for very large $y$ values.
Therefore, $n_{FI}(y)$ has to be $y$-dependent:
\bear\label{18}
&&n_{FI}(y)=\left(h_{F_0}(1)+\frac{y-1}{y_0+1}\right)^{-1}
\nonumber\\
&&y_0+1=\frac{1}{2M_IM_F}(M_I+M_F)^2\ear
$n_{FI}(y)$ is symmetric with respect to $I\leftrightarrow F$
and provides for the correct vector current normalization
at $y=1$. Eq. (\ref{16}) with $M_I^*=M_I+c M_F$ and the form (\ref{18})
chosen for $n_{FI}(y)$ also insure the finiteness of
the form factors in the limit (\ref{15}).

The simple phenomenological model
described here allows us to estimate
the transition from factors occurring in $D$- and $B$-meson decays.
The parameters of the model are the constituent quark masses
$m_i,\ m_f$ and $\varepsilon_{sp}$. The effective spectator mass
$\varepsilon_{sp}$
depends on the initial and final particles. It has to be positive
and smaller than $M_F$ (see eq. (\ref{11})). In case $M_I-m_i\simeq
M_F-m_f$, $\varepsilon_{sp}$ should be equal
to this difference. One may expect this to
occur for the transitions  $B\to D^*$, $B\to D$, and $B\to K^*$.
We will use for these decays
\be\label{19}
M_I-m_i=M_F-m_f=\varepsilon_{sp}=0.32\ {\rm GeV}.\ee
In the more general cases we take for $\varepsilon_{sp}$
the weighted average
\be\label{20}
\varepsilon_{sp}=\frac{m_f}{m_i+m_f}(M_I-m_i)+\frac{m_i}{m_i+m_f}
(M_F-m_f)\ee
which meets the requirements mentioned above. Keeping  $M_I-m_i
\simeq 0.32$ GeV for the initial $D$- or $B$-particles the only
remaining parameter is $m_f<M_F$.
Clearly, the result of the model for the region near $y=1$ is
not fully reliable. For instance, Luke's theorem allows
$1/M^2$ corrections to eq. (\ref{17}) for $I\not=F$ which
in general will reduce the total decay width. An adjustment
of the quark mass parameters to a few well
measured points of the decay spectrum would be very helpful here.

As will be demonstrated below, even without such adjustments
the predictions obtained from this model are in reasonable
agreement with measured form factors as well as with independent
form factor predictions using lattice field theory or QCD sum
rules. Thus, this simple model may be of
practical value until more precise methods become available.
\par
\vspace{.3cm}
\noindent
\begin{center}
\begin{tabular}{|l|c|c|c|c|c|}
\hline
$D^0\to K^*$
&$V$&$A_1$&$A_2$&$\Gamma(K^*)$&$\frac{\Gamma_L(K^*)}{\Gamma_T(K^*)}$\\
\hline\hline
EXP \cite{e}&$1.16\pm0.16$&$0.61\pm0.05$&$0.45\pm0.09$&$5.1\pm0.5$&
$1.15\pm0.17$\\
\hline
SR \cite{a}&$1.1\pm0.25$&$0.50\pm0.15$&$0.60\pm0.15$&$3.8\pm1.5$&
$0.86\pm0.06$\\
LAT \cite{c}&$1.08\pm0.22$&$0.67\pm0.11$&$0.49\pm0.34$&$6.9\pm1.8$&
$1.2\pm0.3$\\
LAT \cite{d}&$1.01^{+0.3}_{-0.13}$&$0.70^{+0.07}_{-0.10}
$&$0.66^{+0.10}_{-0.15}$&$6.0^{+0.8}_{-1.6}$&
$1.06\pm0.16$\\
\hline
model&1.07&0.69&0.73&7.1&0.97\\
\hline
\end{tabular}
\par
\vspace{3mm}
{\bf Table 1}
\setlength{\leftmargin}{-3cm}
Form factors at $q^2=0$ and the decay
widths in $10^{10}\ {\rm sec}^{-1}$\\
for the
semi-leptonic $D^0\to K^*$  transition\\
\end{center}
\par

For the semileptonic decays of $D$-mesons to $K^*$ and $K$-mesons
much experimental information is available. However, the analysis
has been performed assuming pole-type form factors and
not distinguishing the $A_1$ and $F_0$ form factors from the others.
In Table 1 I compare the result of the model with these data
and the results of QCD sum rule estimates and recent lattice
calculations. For the $D\to K^*$ transition
the charm and strange quark masses
$m_c=M_D-0.32$ GeV and $m_f=m_s=0.4$ GeV have been used,
respectively. Considering the small energy release in this decay (and
reducing somewhat the ideal wave function overlap (17)) the
agreement with the data is good apart from a discrepancy concerning 
the $A_2$ form factor. The model may also be applied to
decays involving a $K$- or a $\pi$-meson. Here, however, the result 
depends strongly on the effective quark masses and may be questioned
because of the Goldstone nature of these particles. 
\par
\vspace{.3cm}
\noindent
\hspace{-1cm}
\begin{tabular}{|l|c|c|c|c|c|c|c|}
\hline
$B\to D^*$&
$V$&$A_1$&$A_2$&$V/A_1$&$A_2/A_1$&$\rho^2_{A_1}$&$\Gamma(B\to D^*)$\\
\hline\hline
EXP \cite{f,g}&&&&$1.18\pm0.32$&$0.71\pm0.23$&$0.91\pm0.16$&
$2.9\pm0.2$\\
\hline
SR \cite{mm}&0.58&0.46&0.53&$1.26\pm0.08$&$1.15\pm0.20$&&$(1.7\pm
0.6)10^3|V_{cb}|^2$\\
LAT \cite{d}&&&&&&$0.9^{+2+4}_{-3-2}$&\\
\hline
model&$0.75$&$0.68$&$0.70$&$1.10$&$1.02$&$1.14$&$2.4
\times 10^3|V_{cb}|^2$\\
\hline
\end{tabular}
\newpage
\vspace{-3cm}
\begin{center}
\begin{tabular}{|l|c|c|}
\hline
$B\to D$
&$F_1$&$\Gamma(B\to D)/\Gamma(B\to D^*)$\\
\hline\hline
EXP \cite{f}, \cite{g}&&$0.46\pm0.2$\\
\hline
SR \cite{mm}&$0.62\pm0.06$&$0.53\pm0.11$\\
\hline
model&$0.67$&$0.37$\\
\hline
\end{tabular}
\par
\vspace{0.3cm}
{\bf Table 2}
\setlength{\leftmargin}{-2cm}
Form factors at $q^2=0$, the slope at
$q^2=q^2_{max}$, and the decay
widths in $10^{10}\ {\rm sec}^{-1}$ for the
semi-leptonic $B\to D^*$ and $B\to D$ transitions\\
\end{center}

In Table 2 the model is tested using the more energetic
transitions $B\to D^*e^-\bar\nu_e$ and $B\to De^-\bar\nu_e$.
Here - as mentioned above -  $m_b=M_B-0.32$ GeV, $m_c=M_D^{(*)}-
0.32$ GeV is chosen. 
The parameter $\rho_{A1}^2$ is defined as the negative of the
logarithmic
derivative of $\frac{2}{y+1}A_1(y)$ at $y=1$. The (relative)
amplitudes at $q^2=0$
are not sensitive to the quark mass values. One can
- within the model - assign a 5 \% error only. But the
absolute numbers depend on the normalization
prescription (Eq. (\ref{17})). In Fig. 2 the form factors for 
$B\to D^*$ are plotted as a function of $q^2$.
The differential decay rate - taking  $|V_{cb}|=0.036$ -
is plotted in Fig. 3 together with the CLEO II data points
\cite{g,k}.
For the lifetime of the $B$-meson the value
$\tau_B=1.6\times10^{-12}$ sec is used here and in the following 
tables and figures.
\par
\vspace{.3cm}\noindent
\begin{tabular}{|l|c|c|c|c|c|}
\hline
$B\to K^*$&$T_1(q^2=q^2_{max})$&$T_2(q^2=q^2_{max}$&
$T_1(q^2=0)$&$\Gamma(B\to K^*\gamma)$&$\frac{\Gamma(B\to K^*\gamma)}
{\Gamma(B\to X_s\gamma)}$\\
\hline\hline
EXP \cite{z}&&&&$(2.7\pm0.7)10^{-3}$&$0.18\pm0.07$\\
\hline
SR \cite{w}&$0.80\pm0.06$&$0.28\pm0.04$&$0.38\pm0.06$&&\\
LAT \cite{u}&$1.3\pm0.1$&$0.52\pm0.05$&&&\\
\hline
model&$1.6$&$0.70$&$0.35$&$2.0|V_{st}^*\cdot V_{tb}|^2\ ^\dagger
$&$0.14$\\
\hline
\end{tabular}
\noindent
$^\dagger$ using for the QCD coefficient $C_7$ the value 0.32 and no long-distance contributions.
\par
\begin{center}
\begin{tabular}{|l|c|c|}
\hline
$B\to \rho$&$T_1(q^2=0)$&$\frac{\Gamma(B\to \rho\gamma)}
{\Gamma(B\to K^*\gamma)}$\\
\hline
EXP \cite{z}&&$<0.34$\\
\hline
SR \cite{x}&$0.27\pm0.034$&\\
SR \cite{y}&$0.24\pm0.04$&\\
\hline
model&$0.30$&$0.70|\frac{V_{dt}}{V_{st}}|^2$\\
\hline
\end{tabular}
\par
\vspace{0.3cm}
{\bf Table 3}
\setlength{\leftmargin}{-2cm}
Penguin-induced form factors in $10^{10}\ {\rm sec}^{-1}$
and transition rates for the radiative decays $\bar B\to K^*\gamma$
and $\bar B\to\rho\gamma$\\
\end{center}
In Table 3 the result for the radiative decays $B\to K^*\gamma$
and $B\to\rho\gamma$ are presented choosing for the $B\to\rho$
transition $m_f=m_u=M_\rho/2$.
Besides giving the form
factor $T_1$ at $q^2=0$ we also give
$T_1$ and $T_2$ at $q^2=q^2_{max}$ even though near $y=1$
the calculated amplitudes
depend rather sensitively on the mass parameters.
For $B\to\rho$, $T_1,T_2$ and $T_3$ are plotted in Fig. 4.
(Our definitions of $T_1,T_2,T_3$ (see Appendix A) are such 
that Eq. (4) holds in the heavy quark limit ).
\par
\begin{center}
\begin{tabular}{|l|c|c|c|c|c|c|}
\hline
$\beta\to\rho$
&$V$&$A_1$&$A_2$&$\Gamma(B\to\rho)$&$\frac{\Gamma_L}{\Gamma_T}$\\
\hline\hline
EXP \cite{z}&&&&$1.8\pm0.5$&\\
\hline
SR \cite{ba}&$0.34\pm0.1$&$0.28\pm0.06$&$0.29\pm0.08$&
$(1.4\pm0.4)10^5\cdot |V_{ub}|^2 $&$0.52\pm0.1$\\
LAT \cite{c}&$0.53\pm0.31$&$0.24\pm0.12$&$0.27\pm0.80$&&\\
LAT \cite{s}&$0.37\pm0.11$&$0.22\pm0.05$&$0.49\pm0.22
$&&\\
\hline
model&$0.35$&$0.30$&$0.33$&$1.8\cdot 10^5\cdot |V_{ub}|^2$
&$0.52$\\
\hline
\end{tabular}
\par
\vspace{0.3cm}
{\bf Table 4}
\setlength{\leftmargin}{-2cm}
Form factors at $q^2=0$
and decay widths in $10^8\ {\rm sec}^{-1}$ for the semileptonic
$\bar B\to\rho$  transition\\
\end{center}
\par
In Table 4 the model results for the semileptonic decay
$\bar{B^0}\to\rho^+e^-\bar\nu_e$  are presented and again
compared with typical QCD sum rule and lattice gauge theory
computations. For $B\to\rho$ 
we used $m_f=M_F/2$. The dependence on $m_f$ is such that a smaller value
leads to an increase of the transition rate mainly due to an 
increase of the form factors $V$ and $A_0$ near $q^2 = q^2_{max}$.
In Fig. 5 the form factors for $B\to\rho$ are exhibited.
As first found by P. Ball \cite{bb}, the form factor $A_1$
differs in its $q^2$ behavious from the other form factors
also in heavy-to-light transitions. In quark models this is a
consequence of relativistic covariance \cite{m}.
Fig. 6 shows the
corresponding differential decay width taking
$|V_{ub}|=0.0032$. For the decay $B\to\pi$ we also used $m_f=M_F/2$. 
The model may be 
less suitable for this decay because of the Goldstone nature
of the $\pi$-meson and the strong dependence of the amplitude on $m_f$.
The $E_\pi/m_\pi=y$
distribution of the decay width is shown in Fig. 7.
\par
\subsection*{Summary}
The model presented here may help to understand the
form factors of hadronic currents. Because of its simple analytic form 
detailed predictions for numerous decay processes can easily be 
obtained from it. The model deals in particular with the
dependence of polarization and decay distributions on the quark
mass values.
The entries in Table 4 which refer to the total width for the $B\to\rho$
transition suggest $|V_{ub}|\approx 0.0032$. The decay
distribution shown in Fig 4 has to be checked experimentally, before 
reliable error limits on this number can be given. Also, the 
simple Ansatz for the 
generalized Isgur-Wise function (Eq.(14)) will certainly need 
modifications in the future.
\par
\vspace{.3cm}
\noindent
It is a pleasure to thank Matthias Neubert 
for a helpful discussion.
\subsection*{Appendix A}

\vspace{.5cm}
\renewcommand{\theequation}{A\arabic{equation}}
\setcounter{equation}{0}

For the transition between two pseudoscalar
mesons, $I(p)\to F(p')$,
the weak decay form factors, which parametrize the hadronic
matrix elements of flavour-changing vector
currents are defined
by the formula
\be\label{A.1}
\langle F(p')|V_\mu|I(p)\rangle=\left((p+p')_\mu
-\frac{M_I^2-M_F^2}{q^2}q_\mu\right)
F_1(q^2)+\frac{M_I^2-M_F^2}{q^2}q_\mu F_0(q^2),\ee
where $q_\mu=(p-p')_\mu$ is the momentum transfer.

For the transition of a pseudoscalar into a vector meson,
$I(p)\to F(\eta, p')$, one defines
\bear\label{A.2}
\langle F(\eta,p')|V_\mu|I(p)\rangle&=&\frac{2i}{M_I
+M_F}
\epsilon_{\mu\nu\alpha\beta}\eta^{*\nu}{p'}^\alpha p^\beta V(q^2),
\nonumber\\
\langle F(\eta,p')|A_\mu|I(p)\rangle&=&\left((M_I+M_F)\eta
^{*\mu}A_1(q^2)-\frac{\eta^*\cdot q}{M_I+M_F}
(p+p')_\mu A_2(q^2)\right.\nonumber\\
&&\left.-2M_F\frac{\eta^*\cdot q}{q^2}q_\mu A_3(q^2)\right)
+2M_F\frac{\eta^*\cdot q}{q^2}q_\mu A_0(q^2),\ear
where $\eta_\mu$ is the polarization vector, satisfying $\eta
\cdot p'=0$. Here, the form factor $A_3(q^2)$ is given by the
linear combination
\be\label{A.3}
A_3(q^2)=\frac{M_I+M_F}{2M_F}A_1(q^2)-\frac{M_I-M_F}{2M_F}A_2(q^2).
\ee
The differential decay width for a semileptonic decay
of a pseudoscalar particle to a final pseudoscalar particle
and a massless lepton pair is given by
\be\label{A.4}
\frac{d\Gamma}{dy}=\frac{G_F^2}{12\pi^3}
{M_F^4}{M_I}
(y^2-1)^{3/2}|F_1(y)|^2\ee
The differential transition rate to a vector particle is
\be\label{A.5}
\frac{d\Gamma}{dy}=\frac{G_F^2}{48\pi^3}(y^2-1)^{1/2}\frac{M_F^2}
{M_I}q^2(H^2_++H^2_-+H^2_0)\ee
It contains the helicity amplitudes
\bear\label{A.6}
&&H_0=\frac{M_I^2}{\sqrt{q^2}}\left((y-\frac{M_F}{M_I})
(1+\frac{M_F}{M_I})A_1(y)-2(y^2-1)\frac{M_F}{M_I}\frac{A_2
(y)}{1+\frac{M_F}{M_I}}\right)\nonumber\\
&&H_\pm=M_I\left((1+\frac{M_F}{M_I})A_1(y)\mp(y^2-1)^{1/2}\frac
{2M_F}{M_I}\frac{V(y)}{1+\frac{M_F}{M_I}}\right).\ear

\bigskip
For a transition of a pseudoscalar meson into
a vector meson caused by a Penguin-type process, $I(p)\to
F(\eta,p')$, I define the form factors $T_1,T_2,T_3$
as  follows:
\bear\label{A.7}
&&\langle F(\eta,p')|(\bar  
q_f\sigma_{\mu\nu}q^\nu(1+\gamma_5)q_i)|I(p)\rangle=\nonumber\\
&&\varepsilon_{\mu\nu\alpha\beta}{\eta^*}^\nu_Fp^\alpha
{p'}^\beta 2T_1
(q^2)-\nonumber\\
&&i(\eta^*_\mu(M^2_I-M^2_F)-(\eta^*\cdot q)(p_\mu+p_\mu'))
T_2(q^2)-\nonumber\\
&&i(\eta^*\cdot q)\frac{M_I-M_F}{M_I+M_F}
\left(q_\mu-\frac{q^2}{M_I^2-M^2_F}(p_\mu+p_\mu')\right)
T_3(q^2).\ear
This definition is chosen such that Eq.(4) holds in the heavy
quark limit.

\bigskip
The transition rate for the radiative decay to a vector
particle $I\to F\gamma$ is
\be\label{A.8}
\Gamma=\frac{G_F^2\alpha}{32\pi^4}|V^*_{tf}V_{ti}
|^2C^2_7m^2_iM^3_I\left(1
-\frac{M^2_F}{M^2_I}\right)^3|T_1(q^2=0)|^2.\ee
$C_7$ describes the relevant Wilson coefficient, long-range
contributions are neglected. The formula for the ratio of
the exclusive to the inclusive decay width reads
\be\label{A.9}
\frac{\Gamma(I\to F\gamma)}{\Gamma(I\to X_f\gamma)}
=\left(\frac{M_I}{m_i}\right)^3\left(1-\frac{M_F^2}{M^2_I}\right)
|T_1(q^2=0)|^2.\ee

\subsection*{Appendix B}
\vspace{0.3cm}\noindent
The functions $h_{\cal F}(y)$ \\
\vspace{.3cm}
\renewcommand{\theequation}{B\arabic{equation}}
\setcounter{equation}{0}
\begin{eqnarray}
\lefteqn{h_{F_1} = } \\ \nonumber
& &{1\over {\left( 1 + y  \right) \,\left( 2\,\epsilon_{{\it sp}} + m_{{\it sp}}
  + y \,m_{{\it sp}} \right) \,M_{I}\,M_{F}\,\left( M_{I} + M_{F} \right)
  }}\,(
+ \, 2\,{{\epsilon_{{\it sp}}}^3}\,M_{I} + \epsilon_{{\it sp}}\,m_{f}\,m_{i}\,M_{I} \\
  \nonumber
& &+ \, y \,\epsilon_{{\it sp}}\,m_{f}\,m_{i}\,M_{I} - \epsilon_{{\it
  sp}}\,m_{f}\,m_{{\it sp}}\,M_{I} - y \,\epsilon_{{\it sp}}\,m_{f}\,m_{{\it
  sp}}\,M_{I} - \epsilon_{{\it sp}}\,m_{i}\,m_{{\it sp}}\,M_{I} \\ \nonumber
& &- \,  y \,\epsilon_{{\it sp}}\,m_{i}\,m_{{\it sp}}\,M_{I} + 2\,{{\epsilon_{{\it
  sp}}}^3}\,M_{F} + \epsilon_{{\it sp}}\,m_{f}\,m_{i}\,M_{F} + y \,\epsilon_{{\it
  sp}}\,m_{f}\,m_{i}\,M_{F} \\ \nonumber
& &- \,  \epsilon_{{\it sp}}\,m_{f}\,m_{{\it sp}}\,M_{F} - y \,\epsilon_{{\it
  sp}}\,m_{f}\,m_{{\it sp}}\,M_{F} - \epsilon_{{\it sp}}\,m_{i}\,m_{{\it sp}}\,M_{F}
  - y \,\epsilon_{{\it sp}}\,m_{i}\,m_{{\it sp}}\,M_{F} \\ \nonumber
& &- \,  4\,{{\epsilon_{{\it sp}}}^2}\,M_{I}\,M_{F} - 4\,y \,{{\epsilon_{{\it
  sp}}}^2}\,M_{I}\,M_{F} + m_{f}\,m_{{\it sp}}\,M_{I}\,M_{F} + 2\,y
  \,m_{f}\,m_{{\it sp}}\,M_{I}\,M_{F} \\ \nonumber
& &+ \, {{y }^2}\,m_{f}\,m_{{\it sp}}\,M_{I}\,M_{F} + m_{i}\,m_{{\it
  sp}}\,M_{I}\,M_{F} + 2\,y \,m_{i}\,m_{{\it sp}}\,M_{I}\,M_{F} +
  {{y }^2}\,m_{i}\,m_{{\it sp}}\,M_{I}\,M_{F} \\ \nonumber
& &+ \, \epsilon_{{\it sp}}\,{{M_{I}}^2}\,M_{F} + y \,\epsilon_{{\it
  sp}}\,{{M_{I}}^2}\,M_{F} + \epsilon_{{\it sp}}\,M_{I}\,{{M_{F}}^2} + y
  \,\epsilon_{{\it sp}}\,M_{I}\,{{M_{F}}^2} \,) \\ \nonumber
\end{eqnarray}

\begin{eqnarray}
\lefteqn{h_{F_0} = } \\ \nonumber
& &{1\over {\left( 1 + y  \right) \,\left( 2\,\epsilon_{{\it sp}} + m_{{\it sp}}
  + y \,m_{{\it sp}} \right) \,M_{I}\,\left( M_{I} - M_{F} \right)
  \,M_{F}}}\,(
+ \, 2\,{{\epsilon_{{\it sp}}}^3}\,M_{I} + \epsilon_{{\it sp}}\,m_{f}\,m_{i}\,M_{I} \\
  \nonumber
& &+ \, y \,\epsilon_{{\it sp}}\,m_{f}\,m_{i}\,M_{I} - \epsilon_{{\it
  sp}}\,m_{f}\,m_{{\it sp}}\,M_{I} - y \,\epsilon_{{\it sp}}\,m_{f}\,m_{{\it
  sp}}\,M_{I} - \epsilon_{{\it sp}}\,m_{i}\,m_{{\it sp}}\,M_{I} \\ \nonumber
& &- \,  y \,\epsilon_{{\it sp}}\,m_{i}\,m_{{\it sp}}\,M_{I} - 2\,{{\epsilon_{{\it
  sp}}}^2}\,{{M_{I}}^2} + m_{f}\,m_{{\it sp}}\,{{M_{I}}^2} + y
  \,m_{f}\,m_{{\it sp}}\,{{M_{I}}^2} \\ \nonumber
& &- \,  2\,{{\epsilon_{{\it sp}}}^3}\,M_{F} - \epsilon_{{\it sp}}\,m_{f}\,m_{i}\,M_{F} -
  y \,\epsilon_{{\it sp}}\,m_{f}\,m_{i}\,M_{F} + \epsilon_{{\it sp}}\,m_{f}\,m_{{\it
  sp}}\,M_{F} \\ \nonumber
& &+ \, y \,\epsilon_{{\it sp}}\,m_{f}\,m_{{\it sp}}\,M_{F} + \epsilon_{{\it
  sp}}\,m_{i}\,m_{{\it sp}}\,M_{F} + y \,\epsilon_{{\it sp}}\,m_{i}\,m_{{\it
  sp}}\,M_{F} - y \,m_{f}\,m_{{\it sp}}\,M_{I}\,M_{F} \\ \nonumber
& &- \,  {{y }^2}\,m_{f}\,m_{{\it sp}}\,M_{I}\,M_{F} + y
  \,m_{i}\,m_{{\it sp}}\,M_{I}\,M_{F} + {{y }^2}\,m_{i}\,m_{{\it
  sp}}\,M_{I}\,M_{F} + \epsilon_{{\it sp}}\,{{M_{I}}^2}\,M_{F} \\ \nonumber
& &+ \, y \,\epsilon_{{\it sp}}\,{{M_{I}}^2}\,M_{F} + 2\,{{\epsilon_{{\it
  sp}}}^2}\,{{M_{F}}^2} - m_{i}\,m_{{\it sp}}\,{{M_{F}}^2} - y
  \,m_{i}\,m_{{\it sp}}\,{{M_{F}}^2} - \epsilon_{{\it sp}}\,M_{I}\,{{M_{F}}^2} -
  y \,\epsilon_{{\it sp}}\,M_{I}\,{{M_{F}}^2} \,) \\ \nonumber
\end{eqnarray}

\begin{eqnarray}
\lefteqn{h_{V} = } \\ \nonumber
& &{1\over {\left( 2\,\epsilon_{{\it sp}} + m_{{\it sp}} + y \,m_{{\it sp}}
  \right) \,M_{I}\,M_{F}}}\,(
+ \, \epsilon_{{\it sp}}\,m_{f}\,M_{I} \\ \nonumber
& &- \,  \epsilon_{{\it sp}}\,m_{{\it sp}}\,M_{I} + \epsilon_{{\it sp}}\,m_{i}\,M_{F} -
  \epsilon_{{\it sp}}\,m_{{\it sp}}\,M_{F} + m_{{\it sp}}\,M_{I}\,M_{F} \\ \nonumber
& &+ \, y \,m_{{\it sp}}\,M_{I}\,M_{F} \,) \\ \nonumber
\end{eqnarray}
\begin{eqnarray}
\lefteqn{h_{A1} = } \\ \nonumber
& &{1\over {\left( 1 + y  \right) \,\left( 2\,\epsilon_{{\it sp}} + m_{{\it sp}}
  + y \,m_{{\it sp}} \right) \,M_{I}\,M_{F}}}\,(
- \,  2\,{{\epsilon_{{\it sp}}}^2}\,m_{f} - 2\,{{\epsilon_{{\it sp}}}^2}\,m_{i} \\ \nonumber
& &+ \, 2\,{{\epsilon_{{\it sp}}}^2}\,m_{{\it sp}} + m_{f}\,m_{i}\,m_{{\it sp}} +
  y \,m_{f}\,m_{i}\,m_{{\it sp}} + \epsilon_{{\it sp}}\,m_{f}\,M_{I} \\
  \nonumber
& &+ \, y \,\epsilon_{{\it sp}}\,m_{f}\,M_{I} - \epsilon_{{\it sp}}\,m_{{\it
  sp}}\,M_{I} - y \,\epsilon_{{\it sp}}\,m_{{\it sp}}\,M_{I} + \epsilon_{{\it
  sp}}\,m_{i}\,M_{F} \\ \nonumber
& &+ \, y \,\epsilon_{{\it sp}}\,m_{i}\,M_{F} - \epsilon_{{\it sp}}\,m_{{\it
  sp}}\,M_{F} - y \,\epsilon_{{\it sp}}\,m_{{\it sp}}\,M_{F} + y \,m_{{\it
  sp}}\,M_{I}\,M_{F} \\ \nonumber
& &+ \, {{y }^2}\,m_{{\it sp}}\,M_{I}\,M_{F} \,) \\ \nonumber
\end{eqnarray}

\begin{eqnarray}
\lefteqn{h_{A_2} = } \\ \nonumber
& &{1\over {\left( 1 + y  \right) \,\left( 2\,\epsilon_{{\it sp}} + m_{{\it sp}}
  + y \,m_{{\it sp}} \right) \,{{M_{I}}^2}\,M_{F}}}\,(
- \,  2\,{{\epsilon_{{\it sp}}}^2}\,m_{i}\,M_{I} + 2\,{{\epsilon_{{\it sp}}}^2}\,m_{{\it
  sp}}\,M_{I} \\ \nonumber
& &+ \, \epsilon_{{\it sp}}\,m_{f}\,{{M_{I}}^2} + y \,\epsilon_{{\it
  sp}}\,m_{f}\,{{M_{I}}^2} - \epsilon_{{\it sp}}\,m_{{\it sp}}\,{{M_{I}}^2} - y
  \,\epsilon_{{\it sp}}\,m_{{\it sp}}\,{{M_{I}}^2} \\ \nonumber
& &- \,  2\,{{\epsilon_{{\it sp}}}^2}\,m_{i}\,M_{F} + 2\,{{\epsilon_{{\it sp}}}^2}\,m_{{\it
  sp}}\,M_{F} + \epsilon_{{\it sp}}\,m_{i}\,M_{I}\,M_{F} + y \,\epsilon_{{\it
  sp}}\,m_{i}\,M_{I}\,M_{F} \\ \nonumber
& &- \,  3\,\epsilon_{{\it sp}}\,m_{{\it sp}}\,M_{I}\,M_{F} - 3\,y \,\epsilon_{{\it
  sp}}\,m_{{\it sp}}\,M_{I}\,M_{F} + m_{{\it sp}}\,{{M_{I}}^2}\,M_{F} +
  2\,y \,m_{{\it sp}}\,{{M_{I}}^2}\,M_{F} \\ \nonumber
& &+ \, {{y }^2}\,m_{{\it sp}}\,{{M_{I}}^2}\,M_{F} \,) \\ \nonumber
\end{eqnarray}

\begin{eqnarray}
\lefteqn{h_{A_0} = } \\ \nonumber
& &{1\over {\left( 2\,\epsilon_{{\it sp}} + m_{{\it sp}} + y \,m_{{\it sp}}
  \right) \,M_{I}\,M_{F}\,\left( M_{I} + M_{F} \right) }}\,(
- \,  2\,{{\epsilon_{{\it sp}}}^2}\,m_{f}\,M_{I} + m_{f}\,m_{i}\,m_{{\it sp}}\,M_{I}
  \\ \nonumber
& &+ \, y \,m_{f}\,m_{i}\,m_{{\it sp}}\,M_{I} + \epsilon_{{\it
  sp}}\,m_{f}\,{{M_{I}}^2} + \epsilon_{{\it sp}}\,m_{{\it sp}}\,{{M_{I}}^2} -
  2\,{{\epsilon_{{\it sp}}}^2}\,m_{i}\,M_{F} \\ \nonumber
& &+ \, 2\,{{\epsilon_{{\it sp}}}^2}\,m_{{\it sp}}\,M_{F} + \epsilon_{{\it
  sp}}\,m_{f}\,M_{I}\,M_{F} + \epsilon_{{\it sp}}\,m_{i}\,M_{I}\,M_{F} - 2\,\epsilon_{{\it
  sp}}\,m_{{\it sp}}\,M_{I}\,M_{F} \\ \nonumber
& &- \,  2\,y \,\epsilon_{{\it sp}}\,m_{{\it sp}}\,M_{I}\,M_{F} + \epsilon_{{\it
  sp}}\,m_{i}\,{{M_{F}}^2} - \epsilon_{{\it sp}}\,m_{{\it sp}}\,{{M_{F}}^2} + m_{{\it
  sp}}\,M_{I}\,{{M_{F}}^2} \\ \nonumber
& &+ \, y \,m_{{\it sp}}\,M_{I}\,{{M_{F}}^2} \,) \\ \nonumber
\end{eqnarray}

\begin{eqnarray}
\lefteqn{h_{T_1} = } \\ \nonumber
& &{1\over {\left( 1 + y  \right) \,\left( 2\,\epsilon_{{\it sp}} + m_{{\it sp}}
  + y \,m_{{\it sp}} \right) \,M_{I}\,M_{F}\,\left( M_{I} + M_{F} \right)
  }}\,(
+ \, 2\,{{\epsilon_{{\it sp}}}^3}\,M_{I} + \epsilon_{{\it sp}}\,m_{f}\,m_{i}\,M_{I} \\
  \nonumber
& &+ \, y \,\epsilon_{{\it sp}}\,m_{f}\,m_{i}\,M_{I} - \epsilon_{{\it
  sp}}\,m_{f}\,m_{{\it sp}}\,M_{I} - y \,\epsilon_{{\it sp}}\,m_{f}\,m_{{\it
  sp}}\,M_{I} - \epsilon_{{\it sp}}\,m_{i}\,m_{{\it sp}}\,M_{I} \\ \nonumber
& &- \,  y \,\epsilon_{{\it sp}}\,m_{i}\,m_{{\it sp}}\,M_{I} + 2\,{{\epsilon_{{\it
  sp}}}^3}\,M_{F} + \epsilon_{{\it sp}}\,m_{f}\,m_{i}\,M_{F} + y \,\epsilon_{{\it
  sp}}\,m_{f}\,m_{i}\,M_{F} \\ \nonumber
& &- \,  \epsilon_{{\it sp}}\,m_{f}\,m_{{\it sp}}\,M_{F} - y \,\epsilon_{{\it
  sp}}\,m_{f}\,m_{{\it sp}}\,M_{F} - \epsilon_{{\it sp}}\,m_{i}\,m_{{\it sp}}\,M_{F}
  - y \,\epsilon_{{\it sp}}\,m_{i}\,m_{{\it sp}}\,M_{F} \\ \nonumber
& &- \,  4\,{{\epsilon_{{\it sp}}}^2}\,M_{I}\,M_{F} - 4\,y \,{{\epsilon_{{\it
  sp}}}^2}\,M_{I}\,M_{F} + m_{f}\,m_{{\it sp}}\,M_{I}\,M_{F} + 2\,y
  \,m_{f}\,m_{{\it sp}}\,M_{I}\,M_{F} \\ \nonumber
& &+ \, {{y }^2}\,m_{f}\,m_{{\it sp}}\,M_{I}\,M_{F} + m_{i}\,m_{{\it
  sp}}\,M_{I}\,M_{F} + 2\,y \,m_{i}\,m_{{\it sp}}\,M_{I}\,M_{F} +
  {{y }^2}\,m_{i}\,m_{{\it sp}}\,M_{I}\,M_{F} \\ \nonumber
& &+ \, \epsilon_{{\it sp}}\,{{M_{I}}^2}\,M_{F} + y \,\epsilon_{{\it
  sp}}\,{{M_{I}}^2}\,M_{F} + \epsilon_{{\it sp}}\,M_{I}\,{{M_{F}}^2} + y
  \,\epsilon_{{\it sp}}\,M_{I}\,{{M_{F}}^2} \,) \\ \nonumber
\end{eqnarray}

\begin{eqnarray}
\lefteqn{h_{T_2} = } \\ \nonumber
& &{2\over {\left( 2\,\epsilon_{{\it sp}} + m_{{\it sp}} + y \,m_{{\it sp}}
  \right) \,\left( M_{I} - M_{F} \right) \,{{\left( M_{I} + M_{F} \right)
  }^2}}}\,(
+ \, 2\,{{\epsilon_{{\it sp}}}^3}\,M_{I} + \epsilon_{{\it sp}}\,m_{f}\,m_{i}\,M_{I} \\
  \nonumber
& &+ \, y \,\epsilon_{{\it sp}}\,m_{f}\,m_{i}\,M_{I} - \epsilon_{{\it
  sp}}\,m_{f}\,m_{{\it sp}}\,M_{I} - y \,\epsilon_{{\it sp}}\,m_{f}\,m_{{\it
  sp}}\,M_{I} - \epsilon_{{\it sp}}\,m_{i}\,m_{{\it sp}}\,M_{I} \\ \nonumber
& &- \,  y \,\epsilon_{{\it sp}}\,m_{i}\,m_{{\it sp}}\,M_{I} - 2\,{{\epsilon_{{\it
  sp}}}^2}\,{{M_{I}}^2} + m_{f}\,m_{{\it sp}}\,{{M_{I}}^2} + y
  \,m_{f}\,m_{{\it sp}}\,{{M_{I}}^2} \\ \nonumber
& &- \,  2\,{{\epsilon_{{\it sp}}}^3}\,M_{F} - \epsilon_{{\it sp}}\,m_{f}\,m_{i}\,M_{F} -
  y \,\epsilon_{{\it sp}}\,m_{f}\,m_{i}\,M_{F} + \epsilon_{{\it sp}}\,m_{f}\,m_{{\it
  sp}}\,M_{F} \\ \nonumber
& &+ \, y \,\epsilon_{{\it sp}}\,m_{f}\,m_{{\it sp}}\,M_{F} + \epsilon_{{\it
  sp}}\,m_{i}\,m_{{\it sp}}\,M_{F} + y \,\epsilon_{{\it sp}}\,m_{i}\,m_{{\it
  sp}}\,M_{F} - y \,m_{f}\,m_{{\it sp}}\,M_{I}\,M_{F} \\ \nonumber
& &- \,  {{y }^2}\,m_{f}\,m_{{\it sp}}\,M_{I}\,M_{F} + y
  \,m_{i}\,m_{{\it sp}}\,M_{I}\,M_{F} + {{y }^2}\,m_{i}\,m_{{\it
  sp}}\,M_{I}\,M_{F} + \epsilon_{{\it sp}}\,{{M_{I}}^2}\,M_{F} \\ \nonumber
& &+ \, y \,\epsilon_{{\it sp}}\,{{M_{I}}^2}\,M_{F} + 2\,{{\epsilon_{{\it
  sp}}}^2}\,{{M_{F}}^2} - m_{i}\,m_{{\it sp}}\,{{M_{F}}^2} - y
  \,m_{i}\,m_{{\it sp}}\,{{M_{F}}^2} \\ \nonumber
& &- \,  \epsilon_{{\it sp}}\,M_{I}\,{{M_{F}}^2} - y \,\epsilon_{{\it
  sp}}\,M_{I}\,{{M_{F}}^2} \,) \\ \nonumber
\end{eqnarray}
\begin{eqnarray}
\lefteqn{h_{T_3} = } \\ \nonumber
& &{1\over {\left( 1 + y  \right) \,\left( 2\,\epsilon_{{\it sp}} + m_{{\it sp}}
  + y \,m_{{\it sp}} \right) \,M_{I}\,\left( M_{I} - M_{F} \right)
  \,M_{F}}}\,(
+ \, 2\,{{\epsilon_{{\it sp}}}^3}\,M_{I} + \epsilon_{{\it sp}}\,m_{f}\,m_{i}\,M_{I} \\
  \nonumber
& &+ \, y \,\epsilon_{{\it sp}}\,m_{f}\,m_{i}\,M_{I} - \epsilon_{{\it
  sp}}\,m_{f}\,m_{{\it sp}}\,M_{I} - y \,\epsilon_{{\it sp}}\,m_{f}\,m_{{\it
  sp}}\,M_{I} - \epsilon_{{\it sp}}\,m_{i}\,m_{{\it sp}}\,M_{I} \\ \nonumber
& &- \,  y \,\epsilon_{{\it sp}}\,m_{i}\,m_{{\it sp}}\,M_{I} - 2\,{{\epsilon_{{\it
  sp}}}^2}\,{{M_{I}}^2} - 2\,{{\epsilon_{{\it sp}}}^3}\,M_{F} - \epsilon_{{\it
  sp}}\,m_{f}\,m_{i}\,M_{F} \\ \nonumber
& &- \,  y \,\epsilon_{{\it sp}}\,m_{f}\,m_{i}\,M_{F} + \epsilon_{{\it
  sp}}\,m_{f}\,m_{{\it sp}}\,M_{F} + y \,\epsilon_{{\it sp}}\,m_{f}\,m_{{\it
  sp}}\,M_{F} + \epsilon_{{\it sp}}\,m_{i}\,m_{{\it sp}}\,M_{F} \\ \nonumber
& &+ \, y \,\epsilon_{{\it sp}}\,m_{i}\,m_{{\it sp}}\,M_{F} - m_{f}\,m_{{\it
  sp}}\,M_{I}\,M_{F} - 2\,y \,m_{f}\,m_{{\it sp}}\,M_{I}\,M_{F} -
  {{y }^2}\,m_{f}\,m_{{\it sp}}\,M_{I}\,M_{F} \\ \nonumber
& &+ \, m_{i}\,m_{{\it sp}}\,M_{I}\,M_{F} + 2\,y \,m_{i}\,m_{{\it
  sp}}\,M_{I}\,M_{F} + {{y }^2}\,m_{i}\,m_{{\it sp}}\,M_{I}\,M_{F} +
  \epsilon_{{\it sp}}\,{{M_{I}}^2}\,M_{F} \\ \nonumber
& &+ \, y \,\epsilon_{{\it sp}}\,{{M_{I}}^2}\,M_{F} + 2\,{{\epsilon_{{\it
  sp}}}^2}\,{{M_{F}}^2} - \epsilon_{{\it sp}}\,M_{I}\,{{M_{F}}^2} - y \,\epsilon_{{\it
  sp}}\,M_{I}\,{{M_{F}}^2} \,) \\ \nonumber
\end{eqnarray}

Obviously, one has $h_{T1}(y)=h_{F1}(y)$.
\par
Eqs(B1-B9) are valid for $0^-\to 1^-$ transitions. For $0^-\to 0^-$
transitions $h_{F_1}(y)$ and $h_{F_0}$ have to be multiplied by the
factor $(M_I+M_F)/(m_i+m_f)$ in order to meet requirement (8).
\par

\begin{figure}[htb]
  \begin{center}
    \epsfig{file=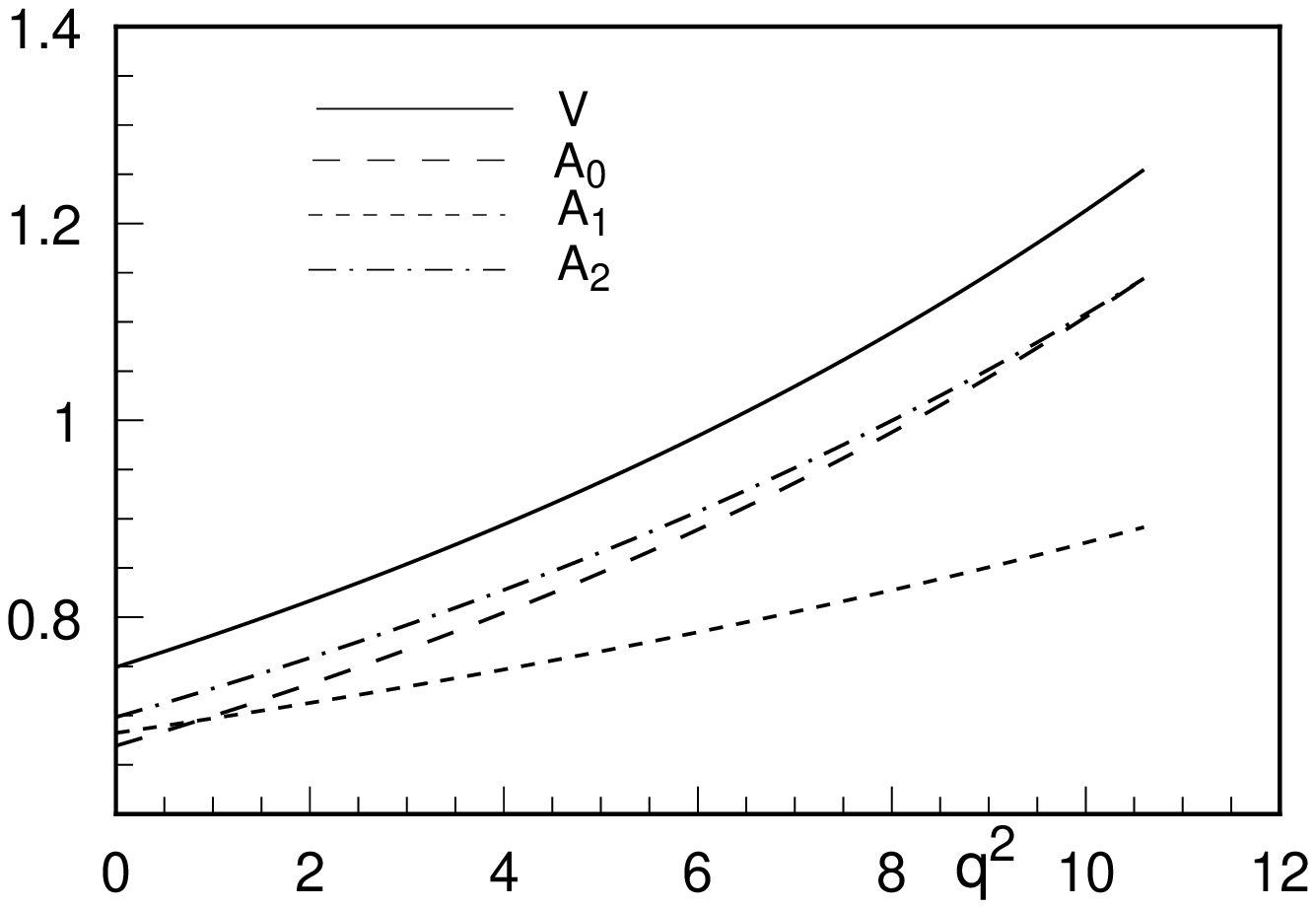,width=.75\hsize}
    \parbox{0.9\textwidth}{\caption{\it{
The $\bar B\to D^*$ form factors
$V,A_0,A_1,A_2$ as a function of $q^2$.}}}
  \end{center}
\end{figure}
\begin{figure}[h]
  \begin{center}
    \epsfig{file=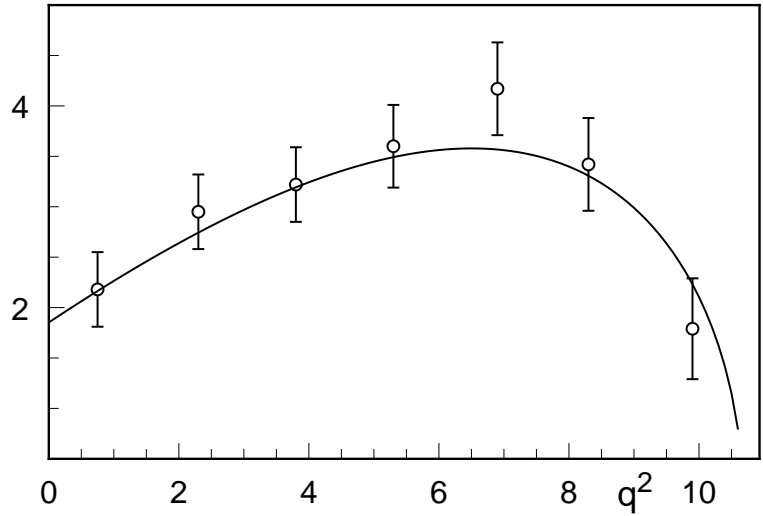,width=.7\hsize}
    \parbox{0.9\textwidth}{\caption{\it{
The differential decay rate in $10^9$ sec$^{-1}$
GeV$^{-2}$ for the semileptonic $\bar B\to D^*$ transition
taking $|V_{cb}|=0.036$. The data points are CLEO II data.}}}
  \end{center}
\end{figure}

\begin{figure}[h]
  \begin{center}
    \epsfig{file=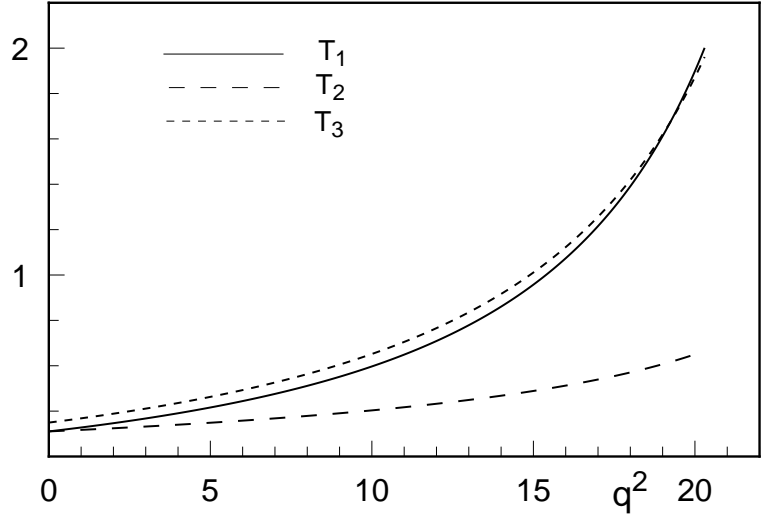,width=.7\hsize}
    \parbox{0.9\textwidth}{\caption{\it{
The form factors $T_1,T_2,T_3$ for the
Penguin-induced $B\to\rho$ transition using $m_f=M_{\rho}/2$.
}}}
  \end{center}
\end{figure}
\begin{figure}[h]
  \begin{center}
    \epsfig{file=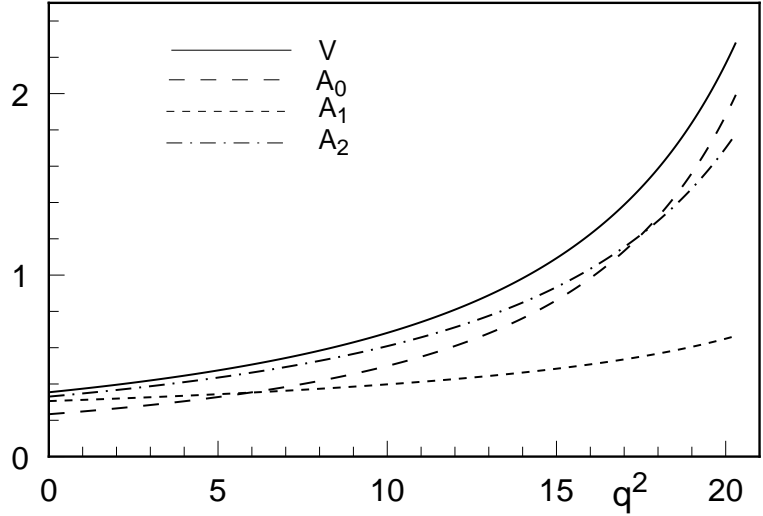,width=.7\hsize}
    \parbox{0.9\textwidth}{\caption{\it{
The $B\to\rho$ form factors $V,A_0,A_1,A_2$ using $m_f=M_{\rho}/2$.}}}
  \end{center}
\end{figure}
\begin{figure}[h]
  \begin{center}
    \epsfig{file=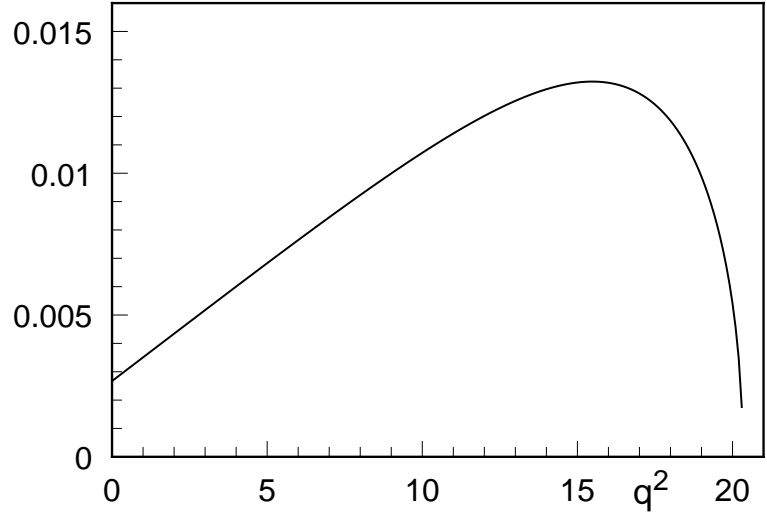,width=.7\hsize}
    \parbox{0.9\textwidth}{\caption{\it{
The differential decay rate in $10^9$ sec$^{-1}$
GeV$^{-2}$ for the semileptonic $\bar B\to\rho$ transition
taking $m_f=M_{\rho}/2$ and $|V_{ub}|=0.0032$.
}}}
  \end{center}
\end{figure}
\begin{figure}[h]
  \begin{center}
    \epsfig{file=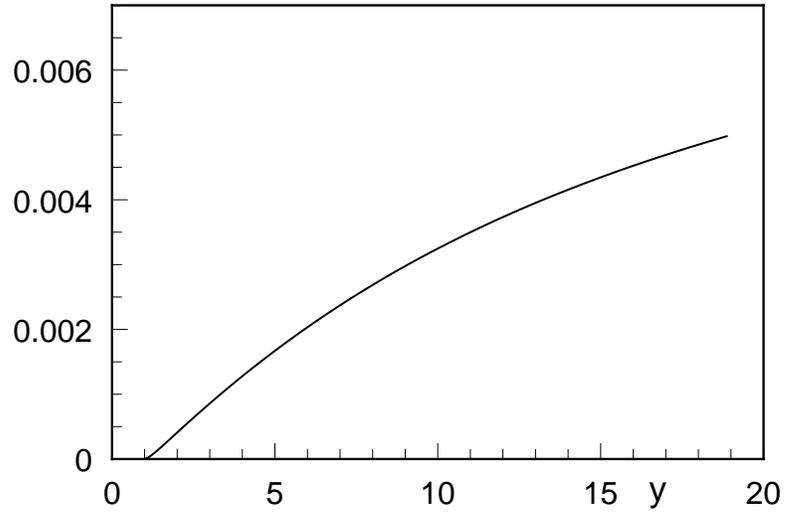,width=.7\hsize}
    \parbox{0.9\textwidth}{\caption{\it{
The differential decay distribution
for the semileptonic $B\to\pi$ transition in $10^9
\ {\rm sec}^{-1}$ as a function of $y$ taking $m_f=M_{\pi}/2$
and $|V_{ub}|=0.0032$.}}}
  \end{center}
\end{figure}
\end{document}